\documentclass[doublecol, figures]{epl2}

\usepackage{amsmath}
\usepackage{amsfonts}
\usepackage{amssymb}

\usepackage{graphicx}

\graphicspath{{figs/}}

\newcommand{\defn}[1]{\emph{#1}}

\newcommand{\defeq}{:=}

\newcommand{\prob}[1]{\mathbb{P}\left[ #1 \right]}

\newcommand{\size}[1]{\left| #1 \right|}

\newcommand{\tauFP}[1]{\tau^{\text{FP}}_{#1}}
\newcommand{\taurec}[1]{\tau^{\text{rec}}_{#1}}

\newcommand{\tautrans}[2]{\tau_{#1\to#2}}	

\newcommand{\tauplus}[1]{\tau^{+}_{#1}}
\newcommand{\tauminus}[1]{\tau^{-}_{#1}}

\newcommand{\ie}{\emph{i.e.}}

\newcommand{\revision}[1]{#1}

\begin{document}

\title{How rare are diffusive rare events?}

\date{\today}

\author{David P.~Sanders
\thanks{E-mail: \email{dps@fciencias.unam.mx}. Current address:
Departamento de F\'isica, Facultad de Ciencias, Universidad Nacional Aut\'onoma
de M\'exico -- Circuito Exterior, Ciudad Universitaria, 04510 M\'exico D.F.,
Mexico.}
\and Hern\'an Larralde}
\institute{Instituto de Ciencias F\'isicas, Universidad Nacional Aut\'onoma de M\'exico --
Apartado Postal 48-3, 62551 Cuernavaca, Morelos, Mexico}

\shortauthor{David P.~Sanders \etal}


\abstract{
We study the time \revision{until} first occurrence, the \emph{first-passage
time},
of rare density fluctuations
in diffusive systems.  We approach the problem using a model consisting of many
independent random walkers on a lattice. The existence of spatial correlations
makes this problem analytically intractable.  However, for a mean-field
approximation in which the walkers can jump anywhere in the system, we obtain a
simple asymptotic form for the mean first-passage time to have a given number
$k$ of particles at a distinguished site. We show numerically, and argue
heuristically, that for large enough $k$, the mean-field results give a good
approximation for first-passage times for systems with nearest-neighbour
dynamics, especially for two and higher spatial dimensions.  Finally, we show
how the results change when density fluctuations anywhere in the system, rather
than at a specific distinguished site, are considered.
}

\maketitle

\section{Introduction}

Rare events control the kinetics of many physical systems. They are
frequently associated with \emph{activated} processes, corresponding
to the crossing of a high free-energy barrier, as happens, for
example, in nucleation processes \cite{DebenedettiBook}. On the other
hand, in diffusive systems, where particles undergo Brownian motion,
there are no such energy barriers.  Here, questions
arise such as: what is the time required for a particle to
first reach a particular region of space, or for a pair of particles
to meet for the first time?  Depending on the circumstances, such
events can be very rare, in which case they may be the limiting step
in the dynamics of the system.  It is then crucial to
understand when such events will occur.

Diffusive systems exhibit universality, in the sense that their
behaviour is often independent of the microscopic details.  It is then
convenient to study simple models, in the hope that the results will
extend to more complicated systems.  A common choice is that of
particles undergoing a \emph{random walk} on a lattice
\cite{WeissBook}.  For a single-particle random walk, first-passage
times to a given site in an infinite system have long been studied
\cite{MontrollWeissRandomWalksFirstPassageJMP1965}, and much is now
known about their statistics \cite{WeissBook, RednerBook}. Recent
progress in this direction was made by Condamin \etal, who studied
first-passage times and distributions for a single particle diffusing
from a source site to a destination site in a confined
region \cite{CondaminFirstPassageRandomWalksBoundedPRL2005,
CondaminRandomWalksFirstPassageConfinedPRE2007,
CondaminFirstPassageScaleInvariantNature2007}.
Other properties, however, are relevant only when many walkers are
present, for example the sequence of times for each to arrive at a
given point \cite{YusteOrderStatsDDimensionalDiffusionsPRE2001}, and
the territory covered after a given number of steps
\cite{LarraldeNumSitesVisitedNWalkersPRA1992,
YusteTerritoryNRandomWalkersPRE1999}.

In this Letter, we study the following first-passage problem for
systems containing many random walkers on a periodic lattice: how long
does it take for $k$ of the walkers to accumulate at a given site?
For $k$ much larger (or much smaller) than the mean density, this
 corresponds to a large local density fluctuation, and is thus a
rare event.  A related quantity was studied numerically in
\cite{ArgyrakisWeissFirstPassageManyWalkersPhysA2006}, as a model of
particles traversing a membrane pore.  Special cases have also been
studied in relation to Ritort's backgammon model
\cite{RitortGlassinessBackgammonModelPRL1995,
GodrecheLuckUrnModelsReviewJPCM2002}, and another related model was
recently used in a study of cooperativity in chemical kinetics
\cite{DOrsognaChouFirstPassageCooperativityQueuingKineticsPRL2005}.
The general problem has, however, received scant attention, despite
its relevance for many physical systems. 

Indeed, the problem pertains
to any systems whose dynamics may be affected by the accumulation of
diffusing particles in a given region.  A particular physical example
that we have in mind is a fluid of hard discs exhibiting \emph{glassy}
behaviour.  When the density of discs is close to
the fluid--solid transition, configurations become ``jammed'',
resulting in slow, glass-like dynamics
\cite{HuertaNaumisRoleRigidityFluidSolidHardDiscsPRL2003}. However,
even in this situation, each disk can be surrounded by a small amount
of ``free volume'' in which the disks ``rattle'' about. This limited
disk motion gives rise to an effective free volume transport, which,
on large scales, may be viewed as diffusive. 
When enough free volume accumulates in a region of the system, a
neighbouring disc can move into the resulting space, causing a large
local rearrangement of the jammed structure.  The dynamics of these
rearrangements is thus controlled by the rare accumulation events,
which can be viewed as density fluctuations of the free volume.

In this work, we focus on the simple multiple random walker problem,
which still poses rather formidable challenges. Actually, for systems
containing many random walkers, exact results are known for
first-passage times from one configuration to another, in terms of
eigenvalues and eigenvectors of the transition matrix, for the case
where a single walker moves at each time step
\cite{KaoMultiboxUrnModelDirectedPRE2003,KaoMultiurn1DRingPRE2004} or
where all walkers can move simultaneously
\cite{NaglerMultiurnModels1DRingPRE2005}.  But these results do not
extend easily to the case of interest here, where we are concerned
with transitions from a large set of rather disparate configurations
(\ie\ all configurations having $k$ walkers at a given site) to
another.

Since diffusive processes give rise to spatial and temporal
correlations that render the problem analytically intractable, we also
consider a mean-field version of this problem, in which particles can
jump not only between neighbouring sites, but to \emph{any} of the
sites in the system with equal probability, \revision{which reduces the
problem} to a
type of Ehrenfest urn model \cite{GodrecheLuckUrnModelsReviewJPCM2002,
AroraExactSolutionsUrnModelsRelaxationPRE1999}.  We show numerically
that this approximation provides a qualitative explanation of the
behaviour of the mean first-passage time also for diffusive dynamics.

\section{The model}

We consider $N$ independent random walkers on a lattice with volume ($=$number of sites)
$V$ and periodic boundary conditions.  In this paper we study
only regular ``cubic'' lattices of \revision{spatial} dimension $d$ (a ring for
$d=1$,
square lattice for $d=2$ and simple cubic lattice for $d=3$).  In 1D, at each
time step a single walker is selected to move right, move left, or
stay where it is, with probabilities $p$, $q$, and $r=1-(p+q)$,
respectively. For 2D square and 3D simple cubic lattices,
we study only the case in which the chosen walker moves one lattice
spacing in any direction with equal probability.  

We are interested in
the event (\ie\ the set of configurations) $S_k$ that a distinguished site, labelled by $0$, is occupied by 
exactly $k$ particles; in particular, we are interested in the \emph{first time} at which this event occurs, that is, when the system configuration lies inside the set $S_k$.
 The mean particle density (number of particles per site) $\rho \defeq N/V$ plays a crucial role here:
for $k$ either much larger
or much smaller than this mean density, the
event $S_k$ is \emph{rare}.  From the static point of view, this means that the probability that the event occurs is small.  From the
dynamical point of view of interest here, it means that the time
needed to actually see the event occur is, on average, very long.

Starting from an initial condition with the particles distributed
randomly, the main question of interest is thus \emph{how
long it takes for the rare event to occur} -- that is, we are
interested in the mean \emph{first-passage time} $\tauFP{k}$ for the
system to reach $S_k$, averaged over all random initial
conditions which are \emph{not in} $S_k$ \cite{CondaminRandomWalksFirstPassageConfinedPRE2007}.
%
In this paper we consider
the case of relatively low mean density, of order $\rho =
1$, and fluctuations to large local densities $k \gg \rho$.  In
systems where the density is high,  a fluctuation in which a site
becomes nearly empty is also a rare event 
with important
physical applications, such as in Ritort's backgammon model
\cite{RitortGlassinessBackgammonModelPRL1995}. This case can also be
studied using the methods described here, as we will discuss
elsewhere.

\section{Mean-field approximation}

In order to make analytical progress, we consider a mean-field version
of the model, in which the walkers may jump to \emph{any} other site,
with equal probability $1/(V-1)$; this is equivalent to
\revision{considering} dynamics on a completely-connected graph.  The model
then reduces to an Ehrenfest \emph{urn model}
\cite{GodrecheLuckUrnModelsReviewJPCM2002}.

In this case, the particles which are not at the distinguished site
`$0$' can be regarded as belonging to a reservoir containing $N-n_0$
particles, the locations of which are irrelevant. \revision{The
occupation number $n_0$ of site `$0$'} then undergoes a Markovian random
walk
whose hopping probabilities depend on its current \revision{position}, for which
exact results are known
\cite{MurthyKehrFirstPassageTimesRandomWalksPRA1989,
PuryCaceresRandomWalksAsymmDisorderedJPA2003}.  In the context of urn
models, further results were found in
\cite{MurthyKehrRelaxationLateStagesBackgammonJPA1997,
AroraExactSolutionsUrnModelsRelaxationPRE1999}.  We approach the
problem from within this context.
  
We denote the mean first-passage time from $S_{l}$ to $S_{m}$ by
$\tautrans{l}{m}$.  It is convenient to first study $\tauplus{k} \defeq
\tautrans{k}{k+1}$, the mean first-passage time from $S_k$ to
$S_{k+1}$.  Suppose that at a given time step, there are $k$ particles
at the distinguished site, so that the system is in $S_k$.  There are
then three possibilities, whose probabilities depend explicitly on
$k$: with probability $\beta_k$, a walker from the reservoir lands at
the distinguished site, thereby reaching $S_{k+1}$; with probability
$\gamma_k$, a walker from the distinguished site leaves to the
reservoir, giving $S_{k-1}$; and with probability $\alpha_k \defeq
1-\beta_k - \gamma_k$, a walker moves around inside the reservoir,
without affecting site $0$, so that the system remains in $S_k$.  In
the mean-field case, these probabilities are independent of the
microscopic configuration, and are given by
\begin{equation}
\beta_k = \frac{N-k}{N}\frac{1}{V-1}(1-r); \qquad \gamma_k =
\frac{k}{N} (1-r).  
\label{eq:defn-of-beta-and-gamma}
\end{equation}
Here, $k/N$ is the probability of choosing a \defn{site} containing $k$ of the $N$ walkers.
The factor $(1-r)$ appears in all such probabilities, and results in a
factor $1/(1-r)$ in the expressions for mean times; for simplicity, we
omit it in the following, considering the case where the particle must
jump, \ie\ $p+q = 1$.

After a given step, which takes time $1$, if the system finds itself in some $S_i$, then the subsequent mean first-passage time
to $S_{k+1}$ is $\tautrans{i}{k+1}$.
By conditioning on the outcome of a step starting at $S_k$, we obtain
\begin{equation} 
 \tauplus{k} 
= \alpha_k (1+\tauplus{k}) + \beta_k (1+\tautrans{k+1}{k+1}) + \gamma_k
(1+\tautrans{k-1}{k+1}).
 \label{eq:basic-tauplus-relation}
\end{equation}
Since only one walker moves at each step, to reach $S_m$ from $S_l$,
we must first pass through all intermediate $S_i$.  We thus have
additivity of mean first-passage times: $\tautrans{l}{m} =
\sum_{i=l}^{m-1} \tauplus{i}$ for $l<m$.  In particular, to reach $S_{k+1}$
from $S_{k-1}$, we must pass through $S_k$, so that
$\tautrans{k-1}{k+1} = \tauplus{k-1} + \tauplus{k}$.  
Furthermore, we have $\tautrans{m}{m} = 0$ for any $m$.
Inserting these results in eq.~\eqref{eq:basic-tauplus-relation}, we get
\begin{equation}
 \tauplus{k} = 1 + \alpha_k \tauplus{k} + \gamma_k (\tauplus{k-1} +
 \tauplus{k}),
\end{equation}
and hence, by rearranging, we obtain a first-order recurrence relation
giving $\tauplus{k}$ in terms of $\tauplus{k-1}$:
\begin{equation} 
 \tauplus{k} = \frac{1}{\beta_k} + \frac{\gamma_k}{\beta_k}
 \tauplus{k-1}; \label{eq:transition-up}
\end{equation}
this result was first derived in
\cite{MurthyKehrFirstPassageTimesRandomWalksPRA1989} by a different
method.  The boundary value $\tauplus{0}$ is the mean of a geometric
distribution with parameter $\alpha_0$, and is hence given by
$\tauplus{0} = \frac{1}{1-\alpha_0} = \frac{1}{\beta_0}$.  By
interchanging $\beta_k$ and $\gamma_k$, we find a similar recurrence
relation in the other direction for the mean time
$\tauminus{k}\defeq\tautrans{k}{k-1}$, obtaining $\tauminus{k} =
(1+\beta_k \tauminus{k+1}) / \gamma_k$. 

An important related concept in random processes is the
\emph{recurrence time} to a set $A$, that is, the time required for
the system to return to $A$, given that it started there \cite{KacRecurrenceThmBullAMS1947}.  The mean
recurrence time $\taurec{k}$ to the set $S_k$ is found, again by
conditioning, to satisfy
\begin{equation} 
 \taurec{k} = 1 + \beta_k \tauminus{k+1} + \gamma_k
 \tauplus{k-1}. \label{eq:transition-recurrence}
\end{equation}

In the situation that we are considering, for fluctuations to $k \gg
\rho$, the mean times $\tauminus{k}$ to decrease $k$ are small, whereas the
times $\tauplus{k}$ to increase it are large.  We thus obtain the
following approximate relation between the recurrence time and the
first-passage time:
\begin{equation}
\tauplus{k-1} \simeq \frac{1}{\gamma_k} \taurec{k} = \frac{N}{k} \taurec{k}.
\end{equation}

The mean first-passage time to the set $S_k$ from a \emph{random} initial
condition, denoted $\tauFP{k}$, is now given by a sum of the $\tautrans{m}{k}$,
each weighted by
the probability of starting in $S_m$.  This is dominated by the
largest term, $\tautrans{0}{k}$, which in turn is dominated by
$\tauplus{k-1}$, \revision{giving}
the approximate result
\begin{equation}
  \tauFP{k} \simeq \tauplus{k-1}  \simeq \frac{1}{\gamma_k} \taurec{k}.
\end{equation}
 
\section{Asymptotics}
The recurrence relations given above yield closed-form results
for mean first-passage times, expressed as sums
\cite{PuryCaceresRandomWalksAsymmDisorderedJPA2003}.  \revision{That}
representation does not, however, allow us to easily understand
\revision{the} behaviour
as a function of the system parameters.  Instead, we focus on
asymptotic results which give the dominant behaviour.  To do so, we
use
the Kac recurrence theorem \cite{KacRecurrenceThmBullAMS1947,
KacProbabilityPhysicalSciencesBook,
CondaminRandomWalksFirstPassageConfinedPRE2007}.
\revision{This states that the}
mean recurrence time to a set $A$ in an ergodic system is
given \revision{exactly} by  
$\taurec{A} = \frac{1}{\prob{A}},$
where $\prob{A}$ is the \revision{stationary probability}
\revision{that the system is in }
$A$.  By using the results of the previous
section, this will then
provide information on mean first-passage times.
\revision{We remark that this
result of Kac is apparently still not well known in the physics literature:
for example,
refs.~\cite{KaoMultiboxUrnModelDirectedPRE2003,KaoMultiurn1DRingPRE2004,
NaglerMultiurnModels1DRingPRE2005}
resorted to much more complicated techniques to calculate Poincar\'e
cycles (\ie\ mean recurrence times) in these systems; see also
\cite{CondaminFirstPassageRandomWalksBoundedPRL2005,
CondaminRandomWalksFirstPassageConfinedPRE2007}.}

We denote the set of microscopic configurations by $\Omega$.  These are
given by vectors of size $N$ containing the positions of all walkers.
Each such configuration is equally likely, since the transition probabilities between
them are symmetric, so that the Kac formula gives the exact mean
recurrence time to the set $S_k$ as
\begin{equation} \label{eq:kac-recurrence-time}
\taurec{k} 
= \frac{1}{\size{S_k}/\size{\Omega}} = \frac{V^N}{ \binom{N}{k} (V-1)^{N-k}},
\end{equation}
where $\size{A}$ denotes the number of configurations in the (finite)
set $A$, obtained by combinatorial arguments, and $\binom{N}{k}$ is a binomial
coefficient.

For large $N$ and $V$, fixed density $\rho = N/V$, and fixed $k\ll N$,
we substitute the asymptotic results $N! / (N-k)! \sim N^k$ and
$1-\frac{1}{V} \sim \exp(-1/V)$ into the exact recurrence time
expression \eqref{eq:kac-recurrence-time}, to obtain the asymptotics
\begin{align}
\taurec{k} 
&\sim k! \left(\frac{V}{N}\right)^k \exp\left(\frac{N-k}{V} \right), 
\end{align}
and hence
\begin{equation}
\taurec{k} \sim k! \, \rho^{-k} \exp(\rho).
\end{equation}
Physically, the total number of particles present should not affect the
dynamics of a given particle, so that the \emph{physical time} is the
number of steps per particle (\ie\ the physical time unit corresponds
to one ``sweep'' through the N particles in the system).  Since
$\tauFP{k} \simeq \frac{N}{k} \taurec{k}$, we finally obtain the
following approximate asymptotic expression for the per-particle
first-passage time to $S_k$:
 \begin{equation}
\frac{1}{N} \tauFP{k} \simeq  (k-1)! \, \rho^{-k} \exp(\rho).
\label{eq:asymptotic-mf-first-passage-time}
\end{equation}
Asymptotically, the mean-field first-passage time thus
depends on $N$ and $V$ only via the density $\rho$ of
walkers for $k$ small compared to the total number of particles.

\begin{figure}[htp]
\includegraphics{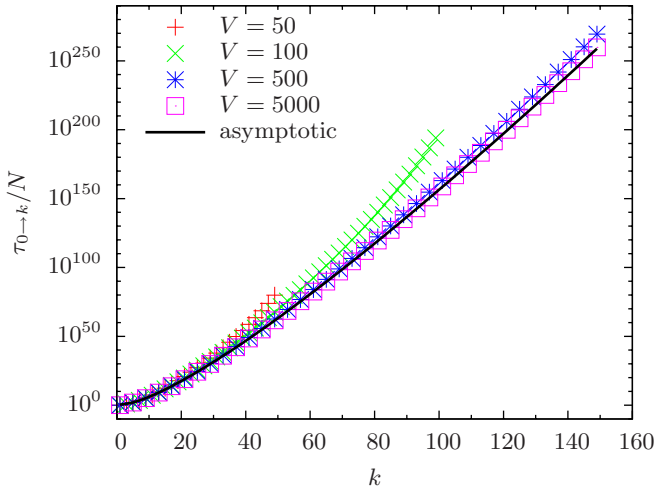}
 \caption{Exact results for mean first-passage
 times $\tautrans{0}{k}$ per particle in mean field for different system sizes $V$, compared to the asymptotic
 result \eqref{eq:asymptotic-mf-first-passage-time}, as a function of
 the size $k$ of the fluctuations.  The mean density is fixed at $\rho=1$.}
 \label{fig:mf-exact-asymp-comparison}
\end{figure}

Figure~\ref{fig:mf-exact-asymp-comparison} compares the exact and
asymptotic mean-field first-passage times per particle (from initial occupation number $0$ to final occupation number $k$) for several
system sizes at fixed density.  The times are indeed independent of
system size, provided that $k$ is not close to $N$.  The curvature visible
in the plot shows that the growth is faster than exponential: indeed,
asymptotically $\log (k!) \sim k \log k - k$, giving a logarithmic
correction on a semi-log plot. For larger values of $k$, a different
approach is needed; however, Arora \etal
\revision{\cite{AroraExactSolutionsUrnModelsRelaxationPRE1999}} showed that
for $k=N$ the asymptotic result is $\tauFP{N} \sim V^N$, which
coincides with the asymptotic behavior of our result when we put
$V=N/\rho$.  

\section{Comparison between mean-field and spatial cases}

The relation between the mean first-passage time $\tauplus{k-1}$ and
the mean recurrence time $\taurec{k}$ was derived in the mean-field
approximation.  Nonetheless, the following argument shows that it
should also provide a reasonable approximation in the case of
diffusive dynamics.  If we start in $S_k$ for a large $k$, then it is
unlikely that we will add another particle to the distinguished site.
Rather, after a waiting time, we will likely drop down to $S_{k-1}$,
and from there most likely to $S_{k-2}$, and so on -- that is, we can
assume that once we have left $S_k$, we escape from there back to a
``random'' condition, from where the system will reach
$S_{k+1}$ via a first-passage process.  Thus, conditioning on whether
we stay in, or leave, $S_k$, we obtain
\begin{align}
\taurec{k}
&\simeq (1 \times \prob{\text{stay}}) +
(\tauFP{\revision{k}} \times \prob{\text{leave}})
\label{eq:simple-relation-rec-FP}\\
&\simeq 1 + \revision{\gamma}_k (\tauFP{k}-1).
\end{align}

Neglecting the fast escape process, we obtain the same approximate relation
$\tauFP{k} \simeq \taurec{k} / \gamma_k$ as before, so that the
mean-field result \revision{should} also describe qualitatively the behaviour in
the
spatial case.  This also explains why higher
dimensions are closer to mean-field: once the system has left $S_k$, it is much harder to return to it, 
since even if we choose a particle in a neighbouring site,
it has many more ways \revision{not to} return (probability $1 - (1/2d)$) than
to return
(probability $1 / 2d$) to the distinguished site.

\begin{figure}[htp]
\includegraphics{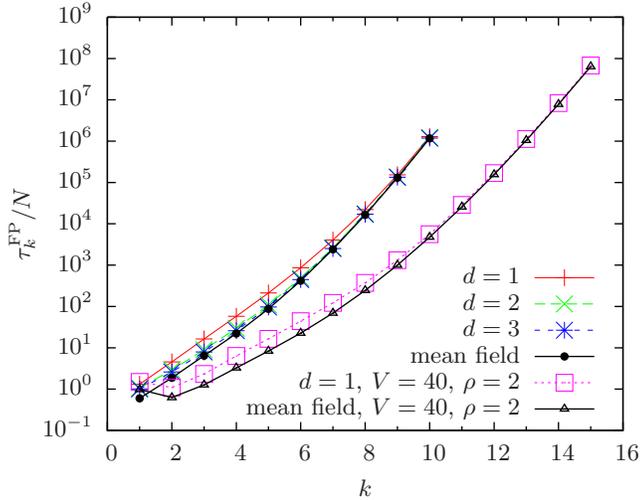}
 \caption{Mean first-passage times $\tauFP{k}$ as a function of $k$ for cubic lattices
 of spatial dimension $d=1,2,3$, for system size $V=729=27^2=9^3$ and density $\rho=1$,  compared to the mean-field result.
 Data are also shown for a 1D diffusive chain with $V=40$, $\rho=2$, and the corresponding mean-field result.  Error bars
 are smaller than the size of the symbols, and lines are shown as a guide for the eye.}
 \label{fig:dimension-comparison}
\end{figure}

Figure~\ref{fig:dimension-comparison} compares the mean-field
results to unbiased nearest-neighbour single-particle motion on
``cubic'' lattices with dimensions $d=1, 2, 3$ and a fixed number of lattice sites.  
We see that the mean-field results reproduce very well the qualitative dependence
of the first-passage time on $k$.   Note that due to the
very rapid increase in the first-passage time as a function of $k$, we
are unable to reach large values of $k$ with brute-force
simulations\footnote{We have explored the possibility of using a
more efficient algorithm along the lines of partial-path sampling
\cite{MoroniBolhuisErpRatesDiffusivePartialPathSamplingJCP2004} to
evaluate these large first-passage times.  Although these methods work very
well for the mean-field case, for diffusive dynamics the strong spatial correlations
present makes their implementation a difficult
task.}.   We also show sample results for a higher density and smaller system size, for which larger values of $k$ can be reached.

\begin{figure}[htp]
 \includegraphics{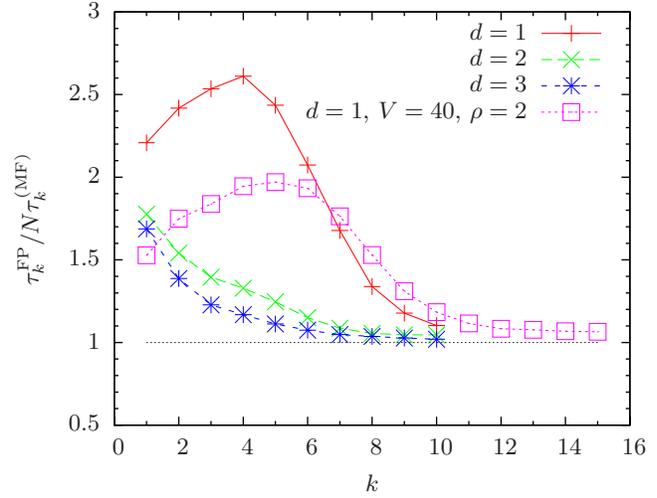}
 \caption{The diffusive data \revision{of} 
fig.~\ref{fig:dimension-comparison},
divided by the corresponding mean-field first-passage times for the same system parameters.}
 \label{fig:dimension-ratio}
\end{figure}

 For a quantitative comparison,  the ratio of the
diffusive to mean-field results is shown in fig.~\ref{fig:dimension-ratio}.
 As expected, mean field is closer
to the spatial results for higher spatial dimension.
\revision{Surprisingly, } for $\revision{d=}2$
spatial dimensions the mean-field result is
also a \emph{quantitatively} good estimate,  provided the event is actually
rare, \ie\ for moderate $k$.  Thus spatial correlation effects are
very strong in 1D, but much less so in higher dimensions, in agreement
with the above discussion. 

Furthermore, in all cases, the mean-field
results become increasingly accurate for larger $k$.  This can be explained as
follows.  For small $k$, in the mean-field case, particles can arrive
rapidly at the distinguished site from any part of the system and
quickly cause the required fluctuation. In the diffusive case,
however, the arrival of particles at the site is controlled by slow
diffusive processes, for which the distance a particle is able to move
in a time $t$ scales only like $\sqrt{t}$.  For large $k$, on the other hand,
the first-passage event takes so long that each particle has enough time to explore
the whole (finite) system by diffusion, and the problem is reasonably
independent of the spatial details.

\revision{This independence can also be explained in terms of
a recent result due to Condamin \etal\
\cite{CondaminFirstPassageScaleInvariantNature2007}.  Our many-walker problem
can be mapped onto the motion of a single walker in an
$(Nd)$-dimensional hypercubic domain
with an appropriate target set.
The result of
ref.~\cite{CondaminFirstPassageScaleInvariantNature2007} shows that
the first-passage time for the applicable case of ``non-compact'' exploration
of the domain should depend rather weakly on the distance to
the target, and hence on whether the dynamics is diffusive or mean-field.
We intend to explore this approach in future work.}

\section{Effect of bias}

\begin{figure}[htp]
\includegraphics{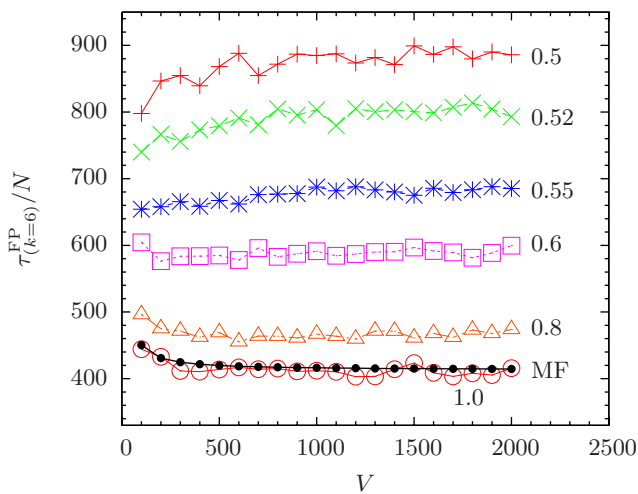}
 \caption{Mean first-passage times per particle to occupation number $k=6$ at a distinguished site, on a 1D lattice as a function of system size $V$, for different values (labelled) of the
 bias parameter $p$, with density $\rho = 1$.  The mean-field (MF) result is shown for comparison (black dots). }
 \label{fig:effect-of-p}
\end{figure}

To further \revision{study} the relation between the spatial and mean-field
cases, we studied in 1D the effect of the bias parameter
$p$, which is the probability that a particle jumps to the right,
always fixing $p+q=1$, so that the chosen particle always jumps.
Figure~\ref{fig:effect-of-p} shows the result of varying $p$ between
$0.5$ (symmetric case with no bias) and $p=1$ (totally asymmetric
motion to the right).  As $p$ varies between these two extremes, the
first-passage time curve interpolates smoothly between the symmetric
and mean-field cases, with the numerical data falling on top of
the mean-field curve for $p=1$.  We can understand this as reflecting the
decreasing influence of spatial correlations.  When $p=1$, spatial
correlations propagate to the right only, so that the incoming
particles to the distinguished site have, in effect, no information
about these spatial correlations, just as in mean field.

\section{Fluctuations anywhere in the system}
Finally, we study the problem of the first-passage time
\revision{$\tilde{\tau}^{\mathrm{FP}}_k$} for
\emph{any} site in the system to reach occupation number $k$.  Even in
mean field, exact calculations for this case are very difficult: we
are aware of solutions only for $V=2$ sites
\cite{MurthyKehrRelaxationLateStagesBackgammonJPA1997} and for the
special case in which $k=N$ \cite{AroraExactSolutionsUrnModelsRelaxationPRE1999}, \ie\ in which all particles collect at
one site, solved using a clever method which
unfortunately does not appear to generalise to other values of $k$.
Solving the case $k=0$ would yield an exact solution for
zero-temperature glassy relaxation in the Ritort backgammon model
\cite{RitortGlassinessBackgammonModelPRL1995,
AroraExactSolutionsUrnModelsRelaxationPRE1999}.

\revision{Instead of an exact solution}, we \revision{can} relate this case to
that
with a distinguished site.  The set of configurations \revision{$\tilde{S}_k$}
forming the rare event is now approximately $V$ times larger than before, since
configurations with $k$ particles on \emph{any} of the $V$ sites are
included in the event.  The Kac method then shows that
recurrence, and hence first-passage, times are approximately $V$ times
\emph{smaller} than before.  Alternatively, we can argue that the
sites are \revision{roughly} independent, with exponentially distributed
first-passage
times.
The minimum of these times --
the time until the first site reaches occupation number $k$ -- is then also
exponentially distributed, with a mean
$V$ times smaller.

\begin{figure}[htp]
\includegraphics{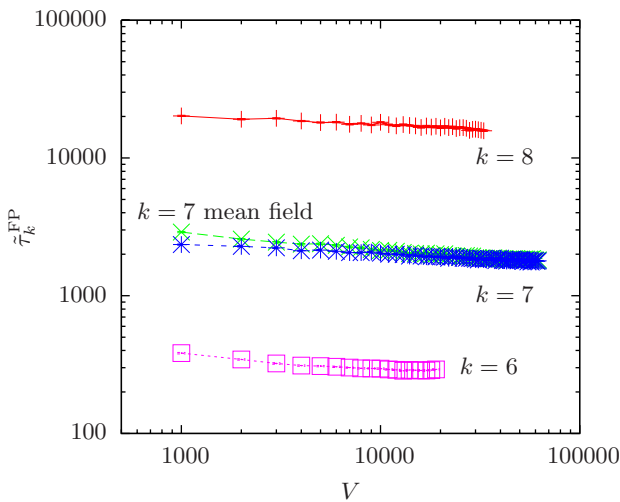}
 \caption{First-passage times \revision{$\tilde{\tau}^{\mathrm{FP}}_k$}, as a
function of system size
 $V$, to reach occupation number $k=6,7,8$
 anywhere in the system, starting from random configurations where no
 sites have this occupation number.  Times shown are numbers of steps.
}
 \label{fig:anywhere}
\end{figure}

To test this, we plot in fig.~\ref{fig:anywhere} the first-passage
time to reach occupation number $k=6,7,8$ somewhere in the system.  In
this figure, the times have \emph{not} been divided by $N$ as in the
previous graphs, and hence they correspond to $\rho V$ times the
physical mean first-passage time to the event.  Since in the
distinguished case, for a fixed density, the times per particle
converge to a constant as $V \to \infty$ (as seen, for example, in
fig.~\ref{fig:effect-of-p}), the above argument would imply that the
raw total number of steps should converge to a constant in the current
case.  In fact, we see that the numerically-determined times decay
slightly \emph{faster} than the $1/V$ expectation, apparently by a
small inverse power of $V$, the exponent of which is roughly
independent of the value of $k$.  (For $k=6$ the times are very short,
and the apparent saturation is not relevant.)

\revision{This observation can be explained as follows. As the system
size $V \to \infty$, configurations with $k$ particles on some site
are much more likely to occur. Each such configuration contributes $0$ to the
calculation of the mean first-passage time. However, the probability of lying
outside the set $\tilde{S}_k$ of such configurations decays exponentially with
$V$. Including them in the calculation thus results in an
exponential decay of the mean first-passage time with $V$, which
merely measures the volume of $\tilde{S}_k$ rather than the dynamics of the
system, and for this reason we disallow such initial conditions.
Nevertheless, as $V$ grows, there is still an ever greater probability
that some sites have occupation numbers \emph{close} to $k$. These
sites can more easily be reached by other particles,
thus lowering the mean first-passage time from its expected $1/V$
behaviour.}

\section{Conclusions}
In summary, we have shown how the mean first-passage time to rare events in
a many-particle random walk model depends on the system parameters.
\revision{We found that the first-passage time to have $k$ particles at a
distinguished
site grows asymptotically as $k!/\rho^k$ for the mean-field case,
and we showed that this agrees}
increasingly well with the \revision{spatial results} for increasingly rare
events.  We also showed how the results change for fluctuations anywhere in the
system.

In the future, we
intend to study
the full probability distribution of
first-passage times.
A further
interesting extension would be to study the statistics of first passage
to large density fluctuations for diffusive processes on heterogeneous
-- \emph{e.g.}\ scale-free -- network structures.

The authors thank George Weiss for posing the problem,
Panos Argyrakis for helpful correspondence, Adri\'an Huerta for
useful discussions, \revision{and the anonymous referees for helpful comments}.
DPS was
supported by the PROFIP program of the Universidad Nacional Aut\'onoma
de M\'exico. Partial financial support from DGAPA-UNAM project IN112307 is also
acknowledged.


\end{document}